\documentclass[prl,twocolumn,amsmath,amssymb,citeautoscript]{revtex4}

\usepackage{amsmath}
\usepackage{amssymb}
\usepackage{epsfig}
\usepackage{graphicx}
\usepackage{wasysym}

\usepackage{amsmath}
\usepackage{amssymb}
\usepackage{epsfig}
\usepackage{graphicx}
\usepackage{wasysym}
\usepackage{bbm}

\newcommand \be{\begin{equation}}
\newcommand \ee{\end{equation}}
\newcommand \bes{\begin{equation*}} 
\newcommand \ees{\end{equation*}}
\newcommand \bea{\begin{eqnarray}}
\newcommand \eea{\end{eqnarray}}
\newcommand \beas{\begin{eqnarray*}} 
\newcommand \eeas{\end{eqnarray*}}
\newcommand \bfg{\begin{figure}}
\newcommand \efg{\end{figure}}
\newcommand \bfgs{\begin{figure*}} 
\newcommand \efgs{\end{figure*}}
\newcommand \bwt{\begin{widetext}}
\newcommand \ewt{\end{widetext}}

\newcommand \fig[1]{Fig.~\ref{#1}}

\begin{document}

\title{Topological insulators in three dimensions from spontaneous symmetry breaking}

\author{Yi Zhang$^{1}$, Ying Ran$^{1,2}$, Ashvin Vishwanath$^{1,2}$}

\affiliation{$^1$Department of Physics, University of California,
Berkeley, CA 94720
\\
$^2$Materials Sciences Division, Lawrence Berkeley National
Laboratory, Berkeley, CA 94720 }
\date{Printed \today}

\begin{abstract}
We study three dimensional systems where strong repulsion leads to
an insulating state via spontaneously generated spin-orbit
interactions. We discuss a microscopic model where the resulting
state is topological. Such topological `Mott'
insulators differ from their band insulator counterparts in that
they possess an additional order parameter, a rotation matrix, that
describes the spontaneous breaking of spin-rotation symmetry. We
show that line defects of this order are associated with protected
one dimensional modes in the {\em strong} topological Mott insulator,
which provides a bulk characterization of this phase.
Possible physical realizations in cold atom
systems are discussed.

\end{abstract}

\maketitle

\section{I. Introduction}

Three dimensional topological insulators (TI) have recently  raised
considerable interest in both theory\cite{key-2,key-8,key-7} and
experiments\cite{key-4,key-3} for their nontrivial topological
properties, which separate them from conventional band
insulators. The hallmark of a TI is its exotic
surface electronic properties. In particular, the strong topological
insulator  has an odd number of Dirac nodes on the surface of
the system, which are stable against moderate perturbations that preserve time
reversal symmetry \cite{key-2}. Such a band structure cannot be
realized in any two dimensional system with time reversal
invariance. There have been experimental realizations of these
predictions in bismuth antimony\cite{key-4,key-4.1,key-4.2} alloys
and in bismuth selenium \cite{key-3, key-3.5}, which have been verified by
angle resolved photoemission spectroscopy.

In two dimensions, an analogous state is the quantum spin hall (QSH)
phase, which has counter-propagating one dimensional edge modes.
Such a state can occur with spin orbit interactions that preserve
spin rotation symmetry about an axis, in which case the oppositely
propagating modes carry opposite quantum numbers for this component
of spin\cite{key-4.4,key-11}. It was shown in Ref.\cite{key-11},
that even in the absence of such spin rotation invariance, the
counter-propagating modes remain protected by time reversal
symmetry. The three dimensional weak topological insulator (WTI) can
be obtained by stacking such 2D QSH states. However, in order to
realize the strong topological insulator, the spin rotation symmetry
must be completely broken.


The TI and QSH phases normally exist in
the systems with strong spin-orbit interaction (SOI) that explicitly
breaks spin rotation symmetry (SRS)\cite{key-10,key-11,key-2}.
However, as pointed out in Ref.\cite{key-1} an extended Hubbard
model on a 2D honeycomb lattice can have spontaneous SRS breaking
and result in a QSH phase, with the right kind of repulsive
interactions\cite{key-1}. Spin rotation symmetry is only preserved
about an axis $\hat{n}$, which is spontaneously chosen, leading to
gapless Goldstone modes. This was termed a topological `Mott'
insulator - the separation of energy scales between the low lying
magnetic excitations and the gapped charge excitations being typical
of Mott insulators. We will also adopt this nomenclature although it
must be noted that local moment physics, often associated with Mott
insulators, does not occur here.
Subsequently, it was argued in \cite{4.5} that
skyrmions of $\hat{n}$ carry charge $2e$.

Here, we consider the analogous problem of a three dimensional system without bare
spin orbit couplings, and full spin rotation symmetry, being driven
into a topological insulator state by strong interactions. The key
difference from the two dimensional case, is that in order to
realize the strong topological insulator, spin rotation symmetry
must be completely broken. Hence the order parameter in this case is
a rotation matrix $\boldsymbol{\overleftrightarrow{R}}\in O(3)$,
similar to superfluid Helium-3 A and B-phases, which leads to a richer set of topological textures. We describe a
microscopic model, an extended Hubbard model on the diamond lattice that, within a mean field
treatment, leads to this phase. The order parameter supports a
number of topological defects. In particular, a vortex like line
defect occurs, but with a $Z_2$ charge. This line defect in the
{\em strong} TI is found to be associated with a pair of gapless fermionic
excitations that travel along its length. These modes are
topologically stable against moderate perturbations such as
impurities and interactions as long as time reversal symmetry is
intact. This is the main result of the paper - an analytical derivation
is provided which relies on the properties of the Dirac equation on a two
dimensional curved surface.

We now contrast our results with other recent work. Similar exotic behavior
also occurs in TIs, along crystal defects such as dislocations. Gapless
fermionic excitations emerge there when  a $Z_{2}$ parameter formed by the
product of the dislocation Burgers vector and three weak topological
insulator (WTI) indices\cite{key-9} is nonzero - which in principle can occur in both the weak and strong TI. In contrast, in the present paper,
the fermionic modes along the line defect are solely determined by the more
elusive {\em strong} index. They are absent in the case of the weak
TI. Thus far, the characterization of the TI phase has relied on the
surface behavior. This result provides a route to identifying the strong topological Mott
insulator via a bulk property.

Similar modes have been identified propagating along a solenoid of $\pi$ flux, inserted into a strong topological insulator \cite{key-9.5}. Here, the $2\pi$ rotation of the electron spin around the line defect leads to a Berry's phase, providing a physical realization of the $\pi$ flux.  Analogous phenomena occur in the context of line defects in superfluid He$_3$-B\cite{key-9.7}.

The order parameter $\boldsymbol{\overleftrightarrow{R}}$  admits a
skyrmion like texture, which is a point object in three dimensions
(Shankar monopole). We find that in contrast to the skyrmion of the
quantum spin Hall effect\cite{4.5}, these are uncharged.

In most solids where electron-electron interactions are important
tend to have some degree of spin-orbit interactions - which will confine the defects. Hence, we
propose realizations of this physics in optical lattices of ultracold atoms, utilizing molecules with multipole moments to obtain the proposed extended Hubbard models. The two dimensional version \cite{key-1} is found to be naturally realized with electric dipoles. Realizing the three dimensional case is more challenging, however  molecules with
electric quadrupole moments confined in optical
lattice can realize some of the key ingredients required.


This paper is organized as the following: In the next section, we
will present the order parameter manifold and the line modes'
$Z_{2}$ dependence on the winding number; in section III, we will
justify our claim with numerical and analytical results; another
texture Shankar monopole will be discussed in section IV; in section
V, we will establish our model Hamiltonian on a diamond lattice and
show the mean field stability of topological Mott insulating (TMI)
phases; we give two possible experimental realizations in cold atom
systems in section VI; we conclude the main result of this paper in
the last section. Hereafter we use $\sigma$ and $\tau$ for the spin
and sublattice degree of freedom, respectively.

\section{II. Topological Mott Insulators and order parameter textures in three dimensions}
In order to describe the Topological Mott insulating phase, we
consider a concrete example in the following. In a subsequent
section, we address the question of how such a phase may be
microscopically realized. To contrast the TMI phase with the regular
topological insulator, consider the model Hamiltonians of a TI
introduced in \cite{key-2}. We consider nearest neighbor hopping
($t_{ij}$) on the sites of a diamond lattice, and spin orbit induced
hopping on next-neighbor sites.
\begin{eqnarray}
\nonumber H_{\rm TI} &=& H_{\rm hop} + H_{\rm SO}\\
\nonumber H_{\rm hop} &=& \underset{\left\langle ij\right\rangle }{\sum
}t_{ij}c_{i}^{\dagger}c_{j}\\
H_{\rm SO} &=&
i\left(8\lambda_{SO}/a^{2}\right)\underset{\left\langle \left\langle
ij\right\rangle \right\rangle
}{\sum}c_{i}^{\dagger}\vec{\sigma}\cdot
\left(\vec{d}_{ij}^{1}\times\vec{d}_{ij}^{2}\right)c_{j}
\end{eqnarray}

where
$c^{\dagger}=\left(c^{\dagger}_{\uparrow},c^{\dagger}_{\downarrow}\right)$
is the electron creation operator and $\vec{\sigma}$ is the spin
pauli matrix. The spin-orbit interaction for a pair of second
neighbor sites $ij$, depend on $d^p_{ij}$ ($p=1,\,2$) the two
nearest neighbor bond vectors connecting the second neighbor sites
$ij$. The spin orbit interactions are thus determined by the crystal
structure. Note, the SU(2) spin rotation symmetry is completely
broken, which is required to realize the strong topological
insulator in three dimensions.

In contrast, in the TMI phase discussed here, the underlying
Hamiltonian possesses full SU(2) spin rotation symmetry that is {\em
spontaneously} broken. The order parameter
$\boldsymbol{\overleftrightarrow{R}}$ then is a rotation matrix that
describes the relative orientation between the real space coordinate
system and the spin axes. The spin orbit term then takes the form:

\begin{equation}
H^{\rm TMI}_{\rm
SO}=i\left(8\lambda_{SO}/a^{2}\right)\underset{\left\langle
\left\langle ij\right\rangle \right\rangle
}{\sum}c_{i}^{\dagger}\vec{\sigma}\cdot\boldsymbol{\overleftrightarrow{R_{l}}}\cdot\left(\vec{d}_{ij}^{1}\times\vec{d}_{ij}^{2}\right)c_{j}
\end{equation}

\begin{figure}
\includegraphics [scale=0.17]{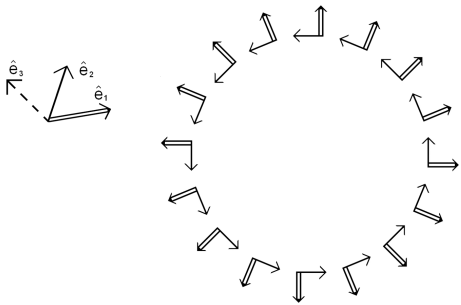}

\caption{An illustration of order parameter $\boldsymbol{R}$ of a
nontrivial line defect in the $x - y$ plane, $\hat{z}$ direction is
translation-invariant. The hollow and solid arrow are the $\hat{x}$
and $\hat{y}$ axes of local coordinate, the $\hat{z}$ axis points
out of the paper for proper rotation.} \label{fig:f03}
\end{figure}

There is one important difference between the TI and TMI: since in
the latter $\boldsymbol{\overleftrightarrow{R}}$ arises from
symmetry breaking it can vary spatially to give rise to a
topologically nontrivial texture.
To identify the topologically stable defects, we first note that the
order parameter manifold is a three dimensional orthogonal matrix
$\boldsymbol{\overleftrightarrow{R}} \in O\left(3\right)$. It can
represent by $\boldsymbol{\overleftrightarrow{R}}=\left(\hat
e_{1},\hat e_{2},\hat e_{3}\right)$, where $\hat e_{i}$ are
orthogonal unit vectors representing the basis vectors of the spin
coordinate system. An example is shown in \fig{fig:f03}.

The target manifold of this order parameter is
$O\left(3\right)=SO\left(3\right)\times Z_{2}$, and $Z_{2}$
determines the chirality
$det\left(\boldsymbol{\overleftrightarrow{R}}\right)=\pm1$, or
whether the rotation is proper or improper. Hereafter we mainly
focus on the continuous $SO\left(3\right)$ part of the order
parameter, for it has nontrivial homotopy groups in two and three
dimensions. Then, each proper rotation can be described by the
parameters $\left(\hat{n},\theta\right)$, where $\hat{n}$ is the
direction of the rotation axis and $\theta\in\left[0,\pi\right]$ is
the rotation angle around it. To visualize it we can map all proper
rotation matrices to a solid ball with radius $\pi$, where $\hat n$
maps to the radial direction and rotation angle maps to the radius
of the image point. Note a rotation of $\pi$ about $\hat n$ is the
same as that of $\pi$ about $-\hat n$, so opposite points on the
spherical surface are identified. The resulting geometry is a three
dimensional projective plane $P^{3}$\cite {key-12}.

We now discuss the topological defects of this order parameter
space. The discrete $Z_2$ symmetry breaking implied by the
$Z_2\times SO(3)$ order parameter leads to domain walls in three
dimensions. More interestingly, line defects also exist (we assume
three spatial dimensions in the following). These can be captured by
considering the order parameter along a closed curve in real space,
which encircles the line defect. This defines a closed loop in the
order parameter space and distinct classes of such closed loops
correspond to the topological line defects. There are two classes of
closed loops for the $SO(3)$ space described above. In addition to
the trivial closed loop, that can be shrunk continuously to a point,
there exists a nontrivial loop that connects the antipodal points
$(\hat{n},\pi)$ and $(-\hat{n},\pi)$. Since these represent the same
rotation, this is in fact a closed loop. Thus there is a nontrivial
line defect characterized just by a $Z_2$ topological charge.
Technically $\pi_{1}\left(SO\left(3\right)\right)=Z_{2}$ \cite
{key-12}. An example is shown in \fig{fig:f03} with translational
invariance along $z$ direction although generically, the line can be
of arbitrary shape and direction. The electronic properties of such
a line defect is studied in the following section - protected one
dimensional modes that propagate along the defect are found in the
case of the strong TMI, but not in the case of the weak TMI. We also
note that since $\pi_{2}\left(SO\left(3\right)\right)=0$, no
nontrivial point defects exist in three dimensions. However, since
$\pi_{3}\left(SO\left(3\right)\right)=Z$, `skyrmion' like textures
(called Shankar monopoles \cite{key-6}) exist in three dimensions.
In contrast to topological defects, they are smooth textures without
a singular core. We investigate the electronic structure of these
objects and find that they are neutral in the large size limit, in
contrast to skyrmions in the two dimensional quantum spin hall state
which carry charge $2e$\cite{4.5}.

\section{III. Electronic Structure of a Line Defect: Numerical and Analytical results}

We study the electronic structure of the $Z_2$ line defect in the
diamond lattice model of a topological Mott insulator discussed
before.
We choose the nearest neighbor hopping in three directions to be
equal $t_{ij}=t>0$ while the fourth is different $t_{ij}=t+\delta
t$. A strong (weak) TI phase occurs if $\delta t> 0$ ($\delta t <
0$) \cite{key-2}. We choose an order parameter texture that
incorporates line defects:

\begin{equation}
H=\underset{\left\langle ij\right\rangle }{\sum
}t_{ij}c_{i}^{\dagger}c_{j}+i\left(8\lambda_{SO}/a^{2}\right)\underset{\left\langle
\left\langle ij\right\rangle \right\rangle
}{\sum}c_{i}^{\dagger}\vec{s}\cdot\boldsymbol{\overleftrightarrow{R_{l}}}\cdot\left(\vec{d}_{ij}^{1}\times\vec{d}_{ij}^{2}\right)c_{j}
\label{H_meanfield}
\end{equation}
with
$\boldsymbol{\overleftrightarrow{R_{l}}}\left(\varphi\right)=\left(\begin{array}{ccc}
cos\left(l\varphi\right) & sin\left(l\varphi\right) & 0\\
-sin\left(l\varphi\right) & cos\left(l\varphi\right)& 0\\
0 & 0 & 1\end{array}\right)$\\

where $\boldsymbol{\overleftrightarrow{R_{l}}}$ depends only on the
azimuthal angle of the atom connecting sites $i$ and $j$. Note, only
the parity of $l$ is topologically stable. Here we study a system of
$24\times24$ with a maximally separated vortex ($l=1$) and
anti-vortex ($l=-1$), and with periodic boundary condition in $x-y$
plane and translational invariance in $z$ direction. Note, time
reversal symmetry is preserved by this Hamiltonian.

\begin{figure}
\includegraphics[scale=0.5]{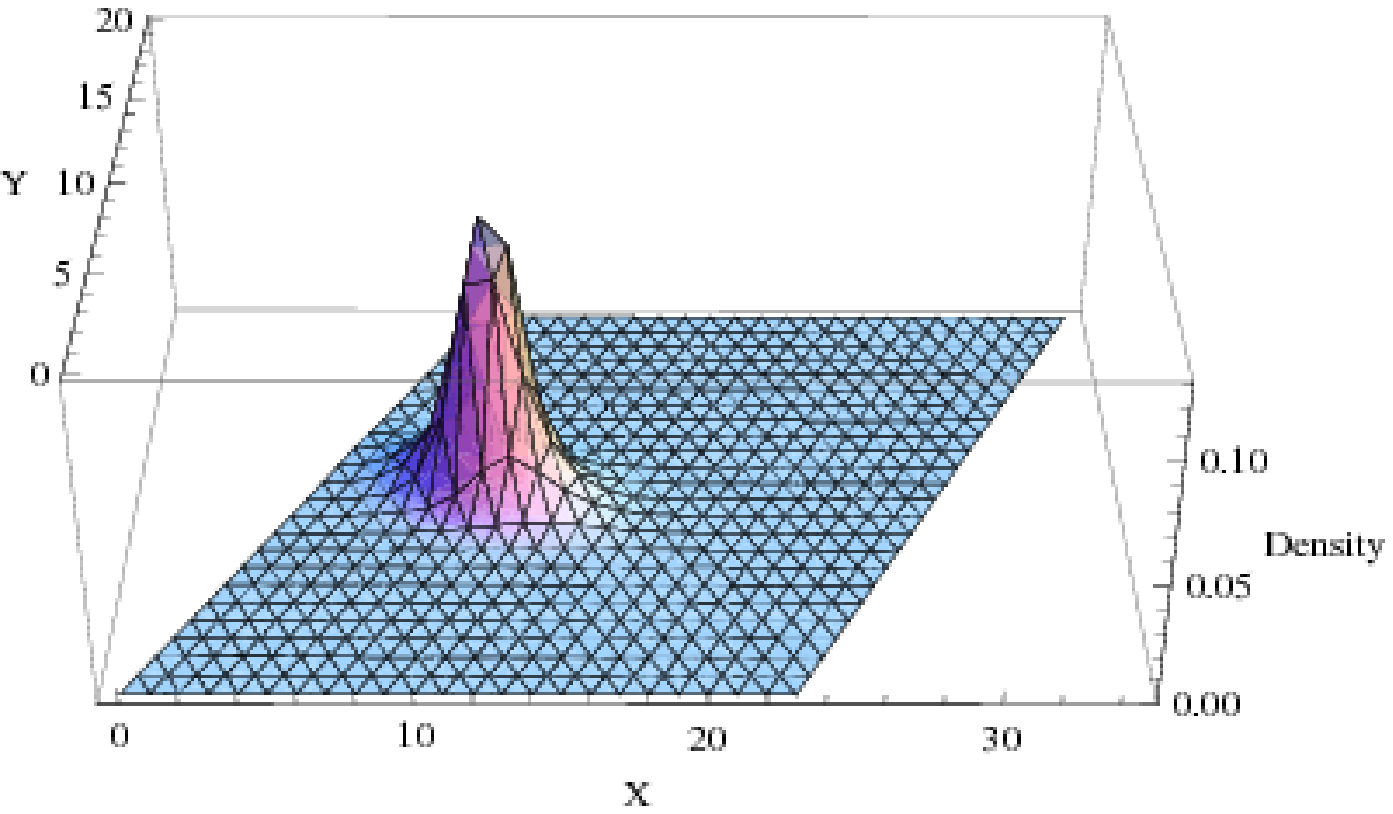}
\\
\includegraphics[scale=0.45]{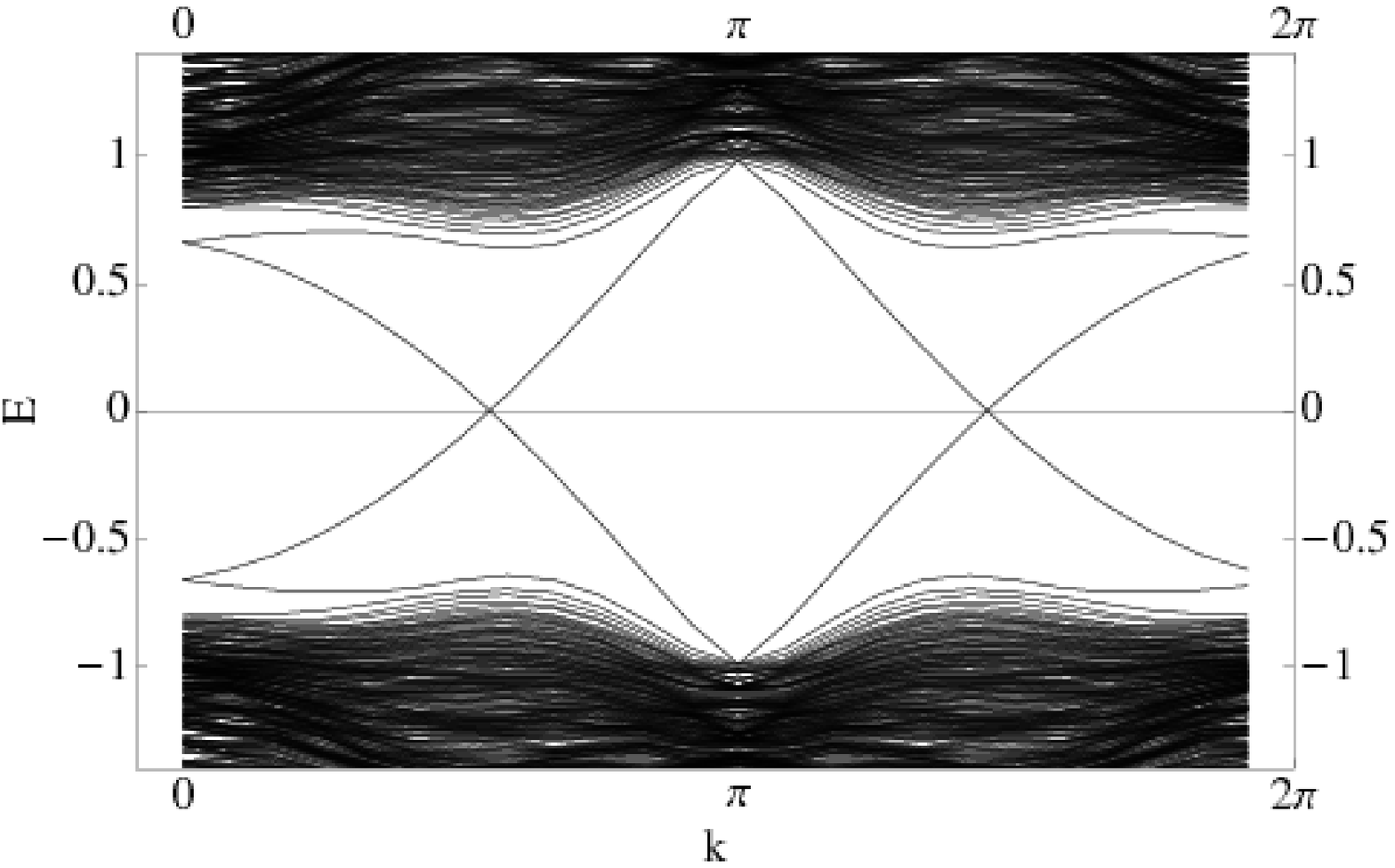}

\caption{Up: The density distribution of a midgap mode
$\left(k=1.05,E=0.25\right)$ in the $\vec{a}_{1},\vec{a}_{2}$ plane,
the mode is well localized at the $l=1$ vortex, the other state
$\left(k=1.05,E=-0.25\right)$ is well localized at the $l=-1$
antivortex; Down: Electronic spectrum of diamond lattice strong TMI
along $p_{z}$ in the presence of a pair of defects. The parameters
used in the Hamiltonian are $t=1.0$, $\delta t=1.0$, and
$\lambda_{SO}=0.125$. } \label{fig:f04}
\end{figure}

For a $\delta t>0$ strong TMI system with $l=\pm 1$, two pairs of
conducting line modes are found in the bulk gap. These states'
density profiles are strongly localized at the cores or the two
defects (\fig{fig:f04} top, shows only one of them for clarity).
Therefore a Kramer pair of modes is localized along the thread of
core. Given the particle hole symmetry that happens to be present in
this model, they cross at zero energy (\fig{fig:f04} bottom). In
contrast, these modes are {\em absent} in the cases of the weak TMI
$\delta t<0$ or if $l=0,\pm2$ in either type of TMI, and the
bandstructure remains fully gapped. This is direct evidence that
these $Z_{2}$ dependent line modes within the bulk identify the
strong TMI.

\begin{figure}
\includegraphics [scale=0.2]{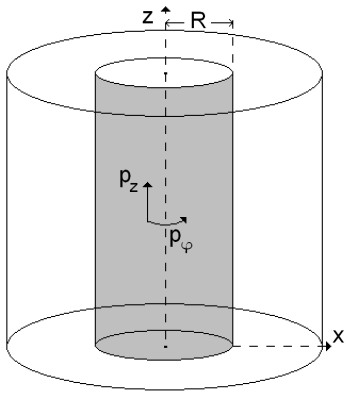}

\caption{Illustration for the model on an effective Hamiltonian on a
curved surface we study. The radius of the cylinder is $R$ with
strong TMI outside and vacuum inside. $p_{z}$ and $p_{\varphi}$ are
momentum along $\hat{z}$ and the azimuthal direction, respectively.
$\theta$ is the angle between $\vec{\sigma}'$ and $\vec{p}$.}
\label{fig:f06}
\end{figure}

An analytical argument for these modes can be developed in several
ways. We can derive these modes based on the 3D Dirac continuum
limit of the Hamiltonian (\ref{H_meanfield}); however, below we
consider deriving these modes using the known properties of surface
states of strong topological insulators. Consider a bulk sample with
a infinitely long cylindrical hole of radius $R$ drilled through its
center. We consider the  low energy states on the cylindrical
surface, both without and with a line defect inserted in the
cylindrical hole \fig{fig:f06}. We show that in the limit of $R
\rightarrow 0$, a propagating midgap mode survives only when the
defect is present; otherwise a fully gapped insulator results.
Interestingly, the key ingredient here is the property of the two
dimensional surface Dirac state, on a curved manifold. In contrast
to the states of a particle on a ring, where the zero angular
momentum eigenstate remains at low energy even when the ring radius
is shrunk to zero, here, the Dirac particle acquires a Berry's phase
of $\pi$ on rotating around the cylinder surface, which excludes the
zero angular momentum state. Hence on shrinking the radius, no low
energy states remain, and the bulk insulator is recovered. However,
if a topological defect of the spin-orbit coupling matrix is
inserted through the cylinder, an additional Berry's phase of $\pi$
is now acquired by the electrons. This ultimately results from the
rotation of the electron's spin by $2\pi$, on following the texture.
With this additional phase, the zero angular momentum mode is
allowed, and a low energy propagating mode results when the cylinder
radius is shrunk to zero. These are the topologically protected one
dimensional modes in the core of the defect. Note, since we are
establishing the presence of a topologically protected excitations,
it is sufficient to use a simple Dirac dispersion for the surface
states of the strong topological insulator, since a general surface
state can always be adiabatically mapped to it. We first describe
the surface Dirac hamiltonian in the presence of curvature, and
apply this to the case of a cylinder with a defect inserted through
it.

{\em Dirac Theory on a Curved Surface:} On a flat surface, spanned
by the unit vectors $\hat{n}_{1},\,\hat{n}_2$, the surface Dirac
Hamiltonian for a strong TI can be taken as:
\begin{equation}
H_{\rm flat} = \sigma_{\tilde{1}}\hat{n}_1 \cdot \vec{p} +
\sigma_{\tilde{2}}\hat{n}_2 \cdot \vec{p}
\end{equation}
 where we assume for simplicity that the spin lies in the same
 plane, with a relative angle of $\theta$ to the momentum, hence $\sigma_{\tilde{1}}=(\hat{n}_1\cos\theta +
 \hat{n}_2\sin\theta)\cdot \vec{\sigma}$ and $\sigma_{\tilde{2}}=(\hat{n}_2\cos\theta
 -\hat{n}_1\sin\theta)\cdot \vec{\sigma}$. Note, for $\theta=0$ this reduces to
 $H=\vec{p}\cdot\vec{\sigma}$, as in \cite{FuKane}. In
 general, the $\sigma$ matrices involve both spin and sublattice
 degrees of freedom, but the essential physics is captured by
 taking them to be simply spin matrices.


On a curved surface there should be additional terms due to the
curvature. The effective Hamiltonian can be systematically derived
\cite{key-14}, as described in the Appendix. Here we just present
the result in the general case when the radii of curvature along the
two tangent directions $\hat{n}_1,\,\hat{n}_2$ are $R_1,\,R_2$
respectively (which are defined via
$\left(\hat{n}_{i}\cdot\vec{p}\right)\hat{n}_{j}=i\hbar\delta_{ij}\hat{n}_1\times
\hat{n}_2/R_{i}$ for $i=1,2$):
\begin{eqnarray}
\nonumber H_{\rm curved} &=& \sigma_{\tilde{1}}\hat{n}_1 \cdot
\vec{p} + \sigma_{\tilde{2}}\hat{n}_2 \cdot \vec{p} \\
 +\frac{\hbar}{2}(\frac1{R_1}&+&\frac1{R_2})\left [ \sin\theta + i \cos \theta (\hat{n}_1\times \hat{n}_2)\cdot \vec{\sigma}\right ]
\end{eqnarray}

We now apply this result to the problem of surface states on the
cylindrical surface of radius $R$ with axis along $\hat{z}$ and
radius $R$ with strong TMI outside and vacuum inside (see
\fig{fig:f06}). We use cylindrical coordinates $z,\,\phi$, hence
$\hat{n}_z = (0,0,1)$ and $\hat{n}_\phi = (-\sin\phi,\cos\phi,0)$,
$\hat{n}_r=-\hat{n}_z\times\hat{n}_\phi$ the two radii of curvature
$R_1=\infty,\,R_2=R$. For simplicity we consider $\theta=0$. The
effective Hamiltonian in the absence of a defect is:
\begin{equation}
H_{0}=\left(\hat{n}_{z}\cdot\vec{\sigma}\right)p_{z}+\left(\hat{n}_{\varphi}\cdot\vec{\sigma}\right)p_{\varphi}+\frac{\hbar}{2R}i\vec{\sigma}\cdot\hat{n}_r
\end{equation}
where $p_z = -i\hbar \partial_z$ and $p_\phi =
-i\frac{\hbar}{R}\partial_\phi$. We can solve for the energies
$H_0\psi = E \psi$ by first performing the unitary transformation
$\psi=U_z(\phi)\psi'$ where $U_z(\phi)=e^{-i\phi\sigma_z/2}$. Note,
since $U_z(\phi+2\pi)=-U_z(\phi)$, the new wavefunctions $\psi'$
satisfy antiperiodic boundary conditions. The transformed
Hamiltonian $H'_0 = p_z \sigma_z + p_\phi \sigma_y$ has eigenvectors
$\psi'=e^{ikz}e^{in\phi}\chi$, where $\chi$ is a fixed spinor. The
energy eigenvalue $E$ then satisfies:
\begin{equation}
E^2_n(k) = \hbar^2(k^2+n^2/R^2) \label{En}
\end{equation}
now, due to the antiperiodicity of the $\psi'$, we require
$n+\frac12$ to be integer. Hence,  $E_n(k)$ in equation \ref{En}
above, all correspond to massive Dirac dispersions whose mass
increases as $R\rightarrow 0$.

We now consider introducing a texture in the order parameter. A
strength `l' is readily introduced by the spin rotation
$U_z(l\phi)$. The Hamiltonian then is:
\begin{equation}
H_l = U_z^\dagger(l\phi) H_0 U_z(l\phi)
\end{equation}
The eigenstates $\psi$ of $H_l$ can be obtained by the unitary
transformation $\psi = U_z([l-1]\phi)\psi'$, and the transformed
Hamiltonian for the wavefunctions $\psi'$ is identical to $H'_0$
above. The energy eigenvalues are then given by Eqn.\ref{En}.
However, the crucial difference is that the single valuedness
property of the wavefunction now requires $n+\frac{l-1}2$ to be an
integer. Hence, for odd values of $l$, when the topologically
nontrivial defect is present, an $n=0$ solution is allowed. The
dispersion of this mode is $E_0 (k)= \pm k$, hence there is a up and
down moving mode, that survives at low energies even when
$R\rightarrow 0$.

The physical picture is: on the surface of a cylinder the momentum
$p_{\varphi}$ is quantized according to boundary condition and the
inter-level spacing is proportional to $R^{-1}$. Also, because of
the spin-momentum relation, when electron circles about the line
defect it picks up a Berry phase of $\left(l+1\right)\pi$ when the
vortex winding number is $l$. When the total phase is an integer
multiple of $2\pi$ it enforces periodic boundary condition,
otherwise anti-periodic boundary condition is applied and the
quantized $p_{\varphi}$ will miss the origin. As we shrink the
cylinder radius to zero, all quantized $p_{\varphi}$ will diverge
except $p_{\varphi}=0$. Therefore, with time reversal symmetry a Kramers pair of line
modes exist when $l$ is odd and its dispersion only depends on
$p_{z}$. Also, the distinction between the strong TMI and weak TMI
can be made clear by the number of surface modes. For a weak TMI,
there are an even number of surface modes, leading to an even number
pairs of line modes with the above reasoning. However, they are
unstable against inter-node scattering. In contrast, a strong TMI
always has an odd number pairs of surface modes.

\section{IV. Shankar monopole in three dimensions}

The line vortex is not the only nontrivial texture in 3D. The other
texture we study is the Shankar monopole, a mapping
$S_{3}\rightarrow SO\left(3\right)$ characterized by the homotopy
classification $\pi_{3}\left(SO\left(3\right)\right)=Z$
\cite{key-6}. Imagine an identical phase faraway from the monopole,
where matrix
$\boldsymbol{\overleftrightarrow{R}}\left(\hat{n},\varphi\right)$ is
independent of $\hat r$, we can identify the infinity of real space
$R_{3}$ thus obtain $S_{3}$. As an specific example here the
$\boldsymbol{\overleftrightarrow{R}}$ matrix connecting $\vec{p}$
and $\vec{\sigma}$ is
$\boldsymbol{\overleftrightarrow{R}}\left(\hat{r},r\right)=exp\left(i\theta\left(r\right)\hat{r}\cdot\hat{J}\right)$
, where $\hat{r}$ and $r$ are the directional vector and distance
from the origin to the site linking $i$ and $j$, and $\hat{J}$ are
classical rotation generators in three dimensions. The rotation axis
$\hat n=\hat r$, and the rotation angle is $\theta\left(r\right)$, a
function $0$ at the origin and smoothly increases to $2l\pi$
$\left(l\in Z\right)$ at infinity. The homotopy class
$\pi_{3}\left(SO\left(3\right)\right)$ is described by integer $l$,
suggesting the base manifold wraps the target manifold $l$ times. It
is protected against any continuous deformation. This is a zero
dimensional defect so the localized states should be localized
charge at the monopole, if any.

However, numerical results show this topological defect does not
carry localized states even in the strong TMI phase. We studied a
single monopole at the center of a $32^3$ unit cell system. To treat
the large system size we found a way to sidestep a complete
diagonalization of the spectrum. Instead, we estimate the boundaries
of the energy eigenstates using ARPACK (Arnoldi Package), and shift
the spectrum so that all states below (above) the band gap are at
negative (positive) energies. Then, we only need to look at the
difference in the number of negative and positive energy
eigenvalues, to determine the monopole charge. This can be done via
an efficient $LDL^T$ factorization, where the Hamiltonian is
factorized into a lower triangular matrix $L$ and a diagonal matrix
$D$. The entries of the diagonal matrix preserves the sign of the
eigenvalues, but not their magnitude. Counting the number of
positive and negative eigenvalues is then readily accomplished.
While smaller system sizes sometimes show charged monopoles, at the
largest sizes, they are found to be neutral. We conclude that the
Shankar monopole texture does not carry charge in the TMI phase.


\section{V. Topological Mott Insulator in a microscopic model - Extended Hubbard Model on the Diamond lattice}
\label{microscopics} 
\begin{figure}
\includegraphics [scale=0.75]{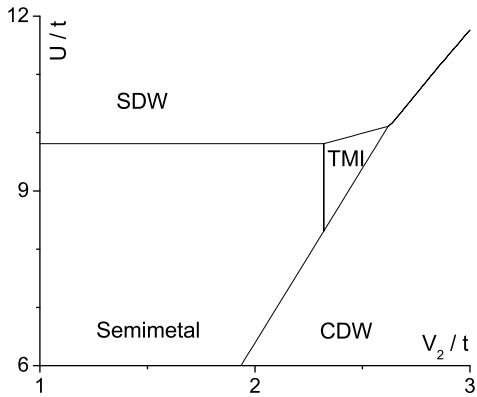}

\caption{Phase diagram for an extended Hubbard model on 3D diamond
lattice. The phase transitions from semimetal to SDW, CDW and TMI
are all second order transitions. Other parameters are
$V_{3}=V_{2}/2$ and $J=0.3t$. The system size is $L=40$ for
calculation.} \label{fig:f02}
\end{figure}

We now discuss the question of realizing the 3D
TMI phase beginning with a microscopic model with full spin rotation
symmetry. We consider an extended Hubbard model on the diamond
lattice within mean field theory. As always, the results of such a
mean field treatment should be treated with caution especially since
strong interactions are involved. Nevertheless, we use this
analytically tractable approach to obtain a range of parameters
where the TMI phase is stabilized over the other obvious candidate
phases - the disordered semi-metal, antiferromagnetic insulator (or
spin density wave: SDW) and the charge density wave (CDW). Realizing
the 3D TMI which completely breaks spin rotation symmetry
spontaneously, requires, in mean field theory, further neighbor
repulsion (between second than third nearest neighbors) as well as a
small antiferromagnetic coupling between second neighbors, as shown
in \fig{fig:f02}.

\begin{figure}
\includegraphics [scale=0.6]{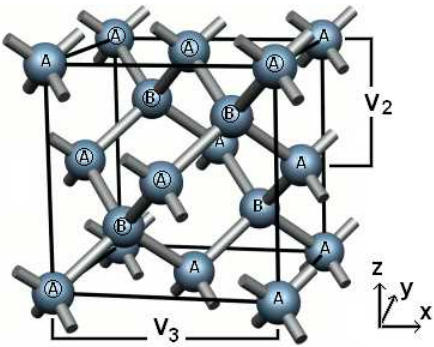}

\caption{(Color Online) A 3D plot of diamond lattice. Each unit cell
contains two sublattices, denoted by $A$ and $B$ respectively, which
each forms an fcc crystal. The repulsive interactions $V_{2}$ and
$V_{3}$ between the second neighbor and third neighbor are shown.
They are both between the atoms of the same sublattice.}
\label{fig:f01}
\end{figure}

We now discuss the details of this mean field treatment.  The model
Hamiltonian we study is an extended Hubbard model on a 3D diamond
lattice at half filling(\fig{fig:f01})
\begin{equation*}
\begin{split}
H=&-\underset{\left\langle ij\right\rangle
,\sigma}{\sum}t\left(c_{i\sigma}^{\dagger}c_{j\sigma}+h.c.\right)+U\underset{i}{\sum}n_{i\uparrow}n_{i\downarrow}
\\&+\underset{ij}{\sum}V_{ij}\rho_{i}\rho_{j}+J\underset{\left\langle \left\langle ij\right\rangle \right\rangle
}{\Sigma}\vec{S}_i\cdot\vec{S}_j
\end{split}
\end{equation*}
\noindent where $t$ is the nearest neighbor hopping strength, $J$ is
the second nearest-neighbor antiferromagnetic coupling strength
between spins $\vec{S}_i= c_{i}^{\dagger}\vec{\sigma}c_{i}$, and
$V_{ij}=V_{2}$ for second nearest-neighbor repulsion, $V_{ij}=V_{3}$
for third nearest-neighbor repulsion while $U$ is the onsite
repulsion strength. All of these operate within the same sublattice
as can be seen from figure \fig{fig:f01}. For simplicity, we assume
no nearest neighbor interaction: $V_{1}=0$; however, as long as
$V_1$ (or fourth-nearest neighbor repulsion $V_{4}$) is not so large
that a nearest-neighbor charge density wave becomes energetically
favored, they can be included but will be irrelevant to our mean
field results. We neglect further neighbor interactions.
$n_{i\sigma}=c_{i\sigma}^{\dagger}c_{i\sigma}$ is the number
operator on site $i$ for spin $\sigma$, and
$\rho_{i}=n_{i\uparrow}+n_{i\downarrow}-1$. Note the Hamiltonian has
full $SU\left(2\right)$ spin rotation symmetry.

Without repulsive interaction the system is in a semi-metal phase
with gapless excitations along lines in the Brillouin zone (`Dirac
lines') and a vanishing DOS at the Fermi level. We turn on
interactions and investigate possible phases including the
Topological Mott Insulator, second and third nearest neighbor charge
density wave (CDW) insulators, and nearest-neighbor spin density
wave(SDW) insulator. Note, the diamond lattice is composed of a FCC
Bravais lattice, plus a two site basis that forms the two
sublattices. The natural SDW phase has alternating spin densities on
two sublattices, resulting from the effective antiferromagnetic
coupling from nearest neighbor hopping and $U$. Note, this is a
$q=0$ order, i.e. preserves lattice translations. In contrast, the
likely CDW orders resulting from second and third neighbor repulsion
break translation symmetry within each sublattice. The phases can be
mapped to a 3D Ising model on an FCC lattice\cite{key-15}, and both
the second and the third nearest neighbor CDWs has the same charge
density distribution for two sublattices but nonuniform distribution
from unit cell to unit cell. Finally, in the TMI phase one develops
second nearest-neighbor correlations $\langle
c^\dagger_{i\sigma}c_{j\sigma'}\rangle \sim
i\vec{\sigma}_{\sigma\sigma'}\cdot\boldsymbol{\overleftrightarrow{R}}\cdot\left(\vec{d}_{i}\times\vec{d}_{j}\right)$
that mimic the spin orbit interaction and breaks the
$SU\left(2\right)$ symmetry completely. Within the mean field
approximation we solve the ground state energy for each phases in
the following, and the resulting phase diagram with fixed $J=0.3t$
is shown in \fig{fig:f02}. Note there is a TMI phase in the center.

We now discuss the mean field energetics of these phases in more
detail.

{\em (i) Semimetal:} For relatively weak interactions, the semimetal
phase arising from the nearest neighbor hopping model on the diamond
lattice remains stable due to the vanishing density of states at the
Fermi energy. This phase retains all symmetries of the Hamiltonian.

{\em (ii) Spin Density Wave} In the limit of large onsite repulsion
$U$ the SDW phase with opposite spin on the two sublattices is
stabilized. More precisely, if $U-24J\gg V_{2}$, the SDW phase
becomes energetically favorable. Define the order parameter $\theta$
\begin{eqnarray*}
\left\langle c_{i\uparrow}^{\dagger}c_{i\uparrow}\right\rangle
=\left\langle c_{j\downarrow}^{\dagger}c_{j\downarrow}\right\rangle
=cos^{2}\theta,\left\langle
c_{i\downarrow}^{\dagger}c_{i\downarrow}\right\rangle =\left\langle
c_{j\uparrow}^{\dagger}c_{j\uparrow}\right\rangle =sin^{2}\theta
\end{eqnarray*}
The ground state energy per unit cell is calculated using mean field
approximation:\begin{equation*}\begin{split}
E_{SDW}=\frac{U}{2}+\left(\frac{U}{2}-12J\right)\chi
-\frac{1}{L^{3}}\underset{k}{\sum}\sqrt{\left(U-24J\right)^{2}\chi+4\left|t\left(k\right)\right|^{2}}
\end{split}\end{equation*}

\noindent Here $\chi=cos^{2}2\theta$. $\chi=0$ denotes the semimetal
phase.
$t\left(k\right)=t\overset{4}{\underset{n=1}{\sum}}e^{i\vec{k}\cdot\vec{t}_{n}}$,
$t_{n}=1,2,3,4$ are the vector from one atom to its nearest
neighbor. $L$ is the number of points along each direction within
the first Brillouin Zone.

{\em (iii) Charge Density Wave} In the limit of strong further
neighbor repulsion $V_2,\,V_3$, a CDW is expected. The problem with
just the density repulsion can be mapped to the FCC lattice Ising
model. In that context it is known that $V_{2}$ will favor a what is
called a Type III CDW phase, while $V_{3}$ will favor a type II CDW
phase\cite{key-15}, described below. The type III CDW phase has Neel
state in the (100) plane and frustrated arrangement between
neighboring planes which leads to a $Z_2$ degeneracy per plane; the
type II CDW phase can be described as a combination of four
independent simple cubic CDWs. The CDW phase is important despite
the frustration. As a matter of fact, in Ref.\cite{key-1} the second
nearest neighbor CDW phase, which the authors neglected, will
dominate the large $V_{2}$ region and the Quantum Anomalous Hall
phase can be realized only with inclusion of $U$. To suppress the
CDW phases, we choose the ratio of $V_{3}/V_{2}$ to be $1/2$ so the
system is close to the transition between type II and type III CDW
phases\cite{key-15}. Hereafter we fix $V_{3}=V_{2}/2$, and point out
that the phase diagram is similar without $V_{3}$  but the TMI phase
will generally occur at a larger $U$ region.

Assuming inversion symmetry and SRS are intact, we define order
parameter $\rho$ for type III CDW phase
\begin{eqnarray*}
\left\langle c_{1\sigma}^{\dagger}c_{1\sigma}\right\rangle
=\left\langle c_{2\sigma}^{\dagger}c_{2\sigma}\right\rangle
=\frac{1+\rho}{2},\left\langle
c_{3\sigma}^{\dagger}c_{3\sigma}\right\rangle =\left\langle
c_{4\sigma}^{\dagger}c_{4\sigma}\right\rangle =\frac{1-\rho}{2}
\end{eqnarray*}
where footnote 1 and 3 are on the first sublattice of two
neighboring unit cells and footnote 2 and 4 are on the second
sublattice and $\sigma$ is the spin index. The energy per original
unit cell is
\begin{equation*}
\begin{split}
&E_{CDW}=3V_{2}\rho^{2}+U\left(1-\rho^{2}\right)/2\\
-&\frac{2}{L^{3}}\underset{k}{\sum}\sqrt{g_{1}^{2}+\left|g_{2}\right|^{2}+\left|g_{3}\right|^{2}\pm\sqrt{4g_{1}^{2}\left|g_{2}\right|^{2}+\left(g_{2}g_{3}^{*}+g_{2}^{*}g_{3}\right)^{2}}}
\end{split}
\end{equation*}

\noindent where $g_{1}=\left(3V_{2}-U/2\right)\rho$,
    $g_{2}=t\left(1+e^{i\vec{k}\cdot\vec{a}_{1}}\right)$,
$g_{3}=t\left(e^{i\vec{k}\cdot\vec{a}_{2}}+e^{i\vec{k}\cdot\vec{a}_{3}}\right)$,
and $\vec{a}_{i}$ are the lattice vectors for the FCC lattice. The
momentum summation is over $L^{2}/2$ points of the unit cell
doubles. The CDW instability is signalled by a non-zero $\rho$.

{\em (iv)  Topological Mott Insulator} More importantly, similar to
the QSH phase in 2D\cite{key-1}, at intermediate couplings
the TMI phase is favored, with order parameters:

\[ \left\langle c_{is}^{\dagger}c_{js'}\right\rangle
=i\left(\vec{\chi}_{ij}\cdot\vec{\sigma}\right)_{ss'}=i\left|\chi\right|\left(\begin{array}{cc}
cos\theta_{ij} & sin\theta_{ij}e^{-i\varphi_{ij}}\\
sin\theta_{ij}e^{i\varphi_{ij}} &
-cos\theta_{ij}\end{array}\right)\]

\noindent where $i,j$ are second nearest neighbors.
Our mean field ansatz assumes for simplicity that all
$\vec{\chi}_{ij}$ have the same magnitude but their  directions are
arbitrary and described by the angles $\theta_{ij}$ and
$\varphi_{ij}$. Note $\vec{\chi}_{ij}=-\vec{\chi}_{ji}$ for
Hermiticity. Lattice translation, rotation and inversion symmetries
are also considered to be intact. This implies that the order
parameters on the other sublattice are the negative of these. We
decouple the Hamiltonian
\begin{equation*}\begin{split}
&H=\frac{UL^{3}}{2}+24L^{3}\left(V_{2}-J\right)\left|\chi\right|^{2}-\underset{k}{\sum}\left(t\left(k\right)c_{k}^{\dagger}I_{\sigma}\otimes\tau^{-}\right.
c_{k}
\\&\left.+h.c.\right)-\underset{k,\vec{d}_{ij}}{\sum}2\left(V_{2}-J\right)\left|\chi\right|sin\left(\vec{k}\cdot\vec{d}_{ij}\right)
\\&\left[cos\theta_{ij}\left(c_{k}^{\dagger}\sigma^{z}\otimes\ \tau_{z}c_{k}\right)+sin\theta_{ij}\left(e^{i\phi_{ij}}c_{k}^{\dagger}\sigma^{+}\otimes\ \tau_{z}c_{k}+h.c.\right)\right]
\end{split}
\end{equation*}
\noindent where the summation is over the two occupied bands.
$\vec{d}_{ij}$ are the vector from on site to its six second nearest
neighbors(one for each opposite pair). We then obtain the ground
state energy per unit cell:\begin{equation*}\begin{split}
&E_{TMI}=\frac{U}{2}+24\left(V_{2}-J\right)\left|\chi\right|^{2}
\\-&\frac{2}{L^{3}}\underset{k}{\sum}\sqrt{\left|t\left(k\right)\right|^{2}+4\left(V_{2}-J\right)^{2}\left|\chi\right|^{2}\left|\underset{\left\langle
\left\langle ij\right\rangle \right\rangle
}{\sum}sin\left(\vec{k}\cdot\vec{d}_{ij}\right)\hat{\chi}_{ij}\right|^{2}}
\end{split}\end{equation*}

\noindent where
$\hat{\chi}_{ij}=\left(sin\theta_{ij}cos\phi_{ij},sin\theta_{ij}sin\phi_{ij},cos\theta_{ij}\right)$.

It is straightforward to see that the energy only depends on
$\left|\chi\right|$ and the relative angles between
$\vec{\chi}_{ij}$. Under global rotation to all $\vec{\chi}_{ij}$
the energy remains invariant and directly leads to an
$SU\left(2\right)$ degeneracy.

This is not a TMI phase in the strict sense, since there are dirac
nodes at the Fermi level. However, an arbitrarily small distortion
of the lattice will introduce anisotropy of nearest neighbor hopping
strengths
$t_{dist}\left(k\right)=\overset{4}{\underset{n=1}{\sum}}t_{n}e^{i\vec{k}\cdot\vec{t}_{n}}$
and an effective mass that leads to a gap of size $\delta
t$\cite{key-2}. In the simplest case that one nearest neighbor
hopping strength is different from the other three
$t_{dist}\left(k\right)=t\left(k\right)+\delta t$, a stronger bond
$\delta t>0$ will lead to a strong TMI phase while a weaker bond
$\delta t<0$ will lead to a weak TMI phase.

{\em Phase Diagram:} For each phase, we search for the global
minimum with respect to its order parameters and compare different
phases. Numerical evaluation of energies was done on a Brillouin
zone with $40\times40\times40$ $k$ space points.  For simplicity we
present the phase diagram with a fixed next nearest
antiferromagnetic coupling strength $J=0.3 t$, and $V_3=0.5V_2$
(\fig{fig:f02}). The semimetal phase exists at small $U$ and $V_{2}$
region; the CDW phase occurs at large $V_{2}$; the SDW phase occurs
at large $U$. Most importantly, there is a TMI phase in the center
of the phase diagram.

This TMI phase has second nearest neighbor correlation similar to
that arising from spin orbit interactions in the Fu-Kane-Mele model
on the same lattice, except, of course for an arbitrary SU(2) spin
rotation
\cite{key-2}, $\hat{\chi}_{ij} \sim
\boldsymbol{\overleftrightarrow{R}}\cdot\left(\vec{d}_{i}\times\vec{d}_{j}\right)$
for each second nearest neighbor pair $i$ and $j$, where
$\vec{d}_{i}$ and $\vec{d}_{j}$ are nearest neighbor bond vectors
connecting this pair of sites, and
$\boldsymbol{\overleftrightarrow{R}}$ is an arbitrary constant three
dimensional rotation matrix.

If we further increase $J$ the stability of the TMI phase is enhanced and it now a wider parameter range.
However, at still larger values, a different TMI phase, that breaks lattice symmetries is realized via a continuous transition. However, since this occurs in the very large $U$ regime, where mean field theory may not be accurate, we do not present further details of this phase.


\section{VI. Towards experimental realizations}


An experimental realization of the TMI phase must contend with two
challenges. First, the system should have weak intrinsic spin-orbit
coupling, but strong interactions. Next, the further neighbour
repulsion should be substantial compared to the nearest neighbor
interactions.
We believe these difficulties can be overcome in cold atom system,
where intrinsic spin orbit couplings are irrelevant, if particles
with electric multipole moments are confined to optical lattice
sites. We first discuss a two dimensional example involving electric
dipoles, for which a fairly definite experimental setup can be
constructed. Although the phase realized here is two dimensional and
does not break spin rotation symmetry completely (a U(1) spin
rotation remains unbroken), it illustrates how the necessary
ingredients can be assembled. Subsequently we discuss ideas for
realizing the three dimensional TMI, the main subject of this paper,
using electric quadrupole moments.


{\em 2D Case: Electric Dipoles on a diamond lattice layer}
Dipole-dipole interactions between hetero-nuclear polar molecules,
such as Rb$_{87}$ and K$_{40}$ have already been shown to be strong
\cite{SilkeOspelkaus}. Consider a fermionic spin 1/2 molecule, with
an electric dipole moment (which is independent of the spin)
confined to the sites of an optical lattice. We note here that the
diamond lattice has a special property that if the dipole-moment is
along the [100] directions, then the nearest neighbor interaction
$V_{1}$ {\em vanishes}. Thus, the second nearest neighbor
interaction $V_{2}$ becomes dominant. However, the difficulty is
that within the twelve second nearest neighbors, only interactions
between neighbors within a plane perpendicular to the dipole moment
are repulsive. This problem can be solved if we restrict the
molecules within a two dimensional (111) layer of the diamond
lattice (still contain both sublattices and essentially two layers
of triangular lattices), as the sites circled in \fig{fig:f01}. Then
if the dipole moment is perpendicular to the plane all possible
nearest neighbor interactions are repulsive.

\begin{figure}
\includegraphics [scale=0.75]{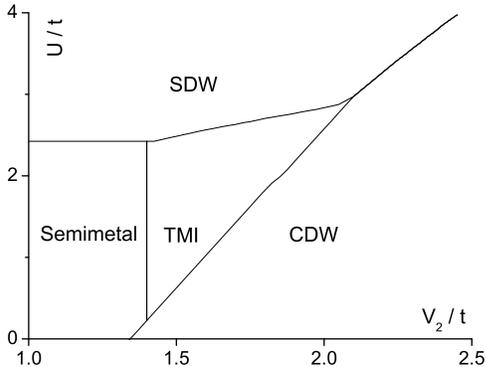}

\caption{Phase diagram for an extended Hubbard model on 2D diamond
lattice layer including both sublattices (essentially honeycomb
lattice). Note its difference from Ref.\cite{key-1}. The TMI is
limited to the center of the phase diagram and only partially breaks
the $SU\left(2\right)$ SRS. The system size is $L=40$ for
calculation.} \label{fig:f05}
\end{figure}

We solve for the mean field phase diagram of this model, as was done
previously for the 3D case. Note, since the lattice is essentially
the honeycomb lattice, this is essentially the model studied in
Ref.\cite{key-1}. There exists a 2D-TMI phase at the center of the
$U - V_{2}$ phase diagram (\fig{fig:f05}). Note this phase diagram
differs from the same model in Ref.\cite{key-1} which has an
extended 2D-TMI phase. This is because we also allow for the second
nearest neighbor CDW that the authors neglected. Though frustrated,
this order will dominate at large $V_{2}$.

The resulting TMI phase in our case has a second nearest neighbor
correlation resembling that of a quantum spin Hall state and SU(2)
spin rotation symmetry is only broken down to U(1). The resulting
order parameter will be $SU\left(2\right)/U\left(1\right)=S_{2}$
instead of $SU\left(2\right)$. Since $\pi_{1}\left(S_{2}\right)=0$
there are no point topological defects - however skyrmions acquire a
charge $2e$ in this phase. Note, the parameters in the phase diagram
seem rather accessible for dipolar molecular systems in this setup,
assuming that the onsite $U$, which results from a combination of
dipolar and microscopic interactions, is not too large.

{\em 3D TMI: Electric quadrupoles on a distorted diamond lattice.} We now discuss some ideas for realizing the 3D TMI. While
these are not as straightforward as the ones discussed earlier, we
nevertheless offer them as one avenue that presents itself at the
current time. Another possible origin of repulsive interaction is
the quadrupole-quadrupole interaction. One can show that for two
parallel uniaxial quadrupoles (quadrupole tensors are identical and
diagonal), the interaction is\begin{equation*}
E=\frac{3Q^{2}}{r^{5}}\left(3-30cos^{2}\theta+35cos^{4}\theta\right)
\end{equation*}
\noindent where $Q$ is the electric quadrupole moment, r is the
distance between two quadrupoles and $\theta$ is the angle between
the quadrupole symmetry axis and the direction between the
quadrupoles.

From this expression if $cos^{2}\theta\simeq 0.742$ the quadrupole
-quadrupole interaction will vanish. Therefore, assume the
quadrupole moment is along the crystal unit cell $\hat{z}$ axis to
retain as many symmetries as possible, if we elongate the lattice
along one direction to make the nearest neighbor satisfy this
condition, we will obtain a system where second nearest neighbor
interaction dominates. This gives $c/a\simeq2.40$. Note the
distortion brings no change to the Hamiltonian we started with.
Within this lattice, the ratio between the repulsive interaction
from the second nearest neighbor out of the quadrupole perpendicular
plane and that from those in plane is $E_{\bot}/E_{\angle}=0.945$.
The $V_{2}$ anisotropy is reasonably small, thus we believe the
physics we discussed with an isotropic $V_{2}$ is unchanged. However,
in this arrangement, $V_3$ is rather small.

To give an reasonable estimate of the quadrupole strength necessary
to drive the system into a TMI phase, we notice the typical nearest
neighbor hopping strength in an cold atom system is of order
$10^{-8}\sim10^{-6}K$ limited by the cooling temperature and the
typical lattice dimension is about the laser wavelength
$0.5\times10^{-6}m$. This leads to a quadrupole of
$10^{-18}\sim10^{-17} e\cdot m^2$, $e$ is unit charge. This is
rather large but maybe realizable in multi-electron molecules. Also,
the critical quadrupole moment can be further reduced by lowering
the temperature or using lasers with shorter wavelength. Finally, a
moderate second nearest neighbor antiferromagnetic coupling may
result from second nearest neighbor hopping super-exchange effect.
We leave for future work construction of a more realistic setting
that can realize the 3D TMI phase.

\section{VII. Conclusion}

We have argued that appropriate repulsive interactions can induce a
spontaneous spin rotation symmetry breaking state, the topological
Mott insulator, where spin orbit couplings are induced by
interactions.

In addition to exotic surface states, line defects of the order
parameter are found to carry protected one dimensional modes along
their length in the strong TMI, which provides a bulk signature of
this phase.
Also, potential experimental directions towards creating these
phases in cold atom system are discussed.  An interesting open
question is whether the form of the spin-orbit interactions near a
line defect, and hence these protected line modes, can be realized
by suitably modifying the atomic structure in a strong topological
band insulator.

\section{VII. Acknowledgements}
We thank T. Senthil, Ari Turner and D. H. Lee for discussions, and acknowledge support from NSF DMR-0645691.
\

\section{Appendix: Effective Hamiltonian on a curved TI surface}

It is known that the effective Hamiltonian for a TI surface has the
form of a Dirac equation: $H=\vec{p}\cdot\vec{\sigma}$ when the spin
and momentum are parallel and
$H=\hat{n}\cdot\left(\vec{p}\times\vec{\sigma}\right)$ when they are
perpendicular, where $\hat{n}$ is the normal direction of the
surface. Both the momentum and spin are confined to the surface.
More generally, when the angle between spin and momentum is $\theta$
the effective Hamiltonian is:\begin{equation*}
\begin{split}
H&=\left(\hat{n}_{1}\cdot\vec{\sigma}cos\theta+\hat{n}_{2}\cdot\vec{\sigma}sin\theta\right)\left(\hat{n}_{1}\cdot\vec{p}\right)
\\&+\left(\hat{n}_{2}\cdot\vec{\sigma}cos\theta-\hat{n}_{1}\cdot\vec{\sigma}sin\theta\right)\left(\hat{n}_{2}\cdot\vec{p}\right)
\end{split}\end{equation*}

\noindent where $\hat{n}_{1}$ and $\hat{n}_{2}$ are orthogonal
directions in plane.

One would tends to apply the same Hamiltonian to a curved surface.
However, here we claim the effective Hamiltonian on an arbitrary
shaped TI surface is:\begin{equation*}
\begin{split}
H&=\left(\hat{n}_{1}\cdot\vec{\sigma}cos\theta+\hat{n}_{2}\cdot\vec{\sigma}sin\theta\right)\left(\hat{n}_{1}\cdot\vec{p}\right)
\\&+\left(\hat{n}_{2}\cdot\vec{\sigma}cos\theta-\hat{n}_{1}\cdot\vec{\sigma}sin\theta\right)\left(\hat{n}_{2}\cdot\vec{p}\right)
\\&+\frac{\hbar}{2}\left(\frac{1}{R_{1}}+\frac{1}{R_{2}}\right)sin\theta+\frac{i\hbar}{2}\left(\vec{\sigma}\cdot\hat{n}_{3}\right)\left(\frac{1}{R_{1}}+\frac{1}{R_{2}}\right)cos\theta
\end{split}\end{equation*}
\noindent where $\hat{n}_{1}$ and $\hat{n}_{2}$ are tangent vectors
of the surface with radius of curvature $R_{1}$ and $R_{2}$,
respectively, and $\hat{n}_{3}$ is the normal vectors. $\theta$ is
the constant angle between spin and momentum in the corresponding
flat surface effective theory. Note when the surface is flat, the
last two terms vanish and the Hamiltonian goes back to the one
describing the flat surface mode.

This Hamiltonian can be systematically derived from the inclusion of
the connection form for a curved space\cite{key-12, key-14}, but an
alternative method we used is to ensure hermiticity and
anticommutation relation $\{H,\vec{\sigma}\cdot\hat{n}_{3}\}=0$
since the spin is in the surface plane\cite{Ari}, with the help of
the relations:
$\left(\hat{n}_{i}\cdot\vec{p}\right)\hat{n}_{j}=i\hbar\delta_{ij}\hat{n}_{3}/R_{i}$,
$\left(\hat{n}_{i}\cdot\vec{p}\right)\hat{n}_{3}=-i\hbar\hat{n}_{i}/R_{i}$
for $i=1,2$, and
$\left(\vec{\sigma}\cdot\hat{A}\right)\left(\vec{\sigma}\cdot\hat{B}\right)=\hat{A}\cdot\hat{B}+i\hat{\sigma}\cdot\left(\hat{A}\times\hat{B}\right)$.

The additional terms arising from the space curvature are canceled
by the inclusion of the connection form. Thus, the above effective
Hamiltonian well describes a curved TI surface.

\end{document}